\documentstyle[floats,prl,aps,epsf,preprint]{revtex}

\begin{document}
\bibliographystyle{plain}
\title{Hawking Radiation of the Brane-Localized Graviton from a 
$(4+n)$-dimensional Black Hole} 
\author{ 
D. K. Park\footnote{Email:dkpark@hep.kyungnam.ac.kr 
}}
\address{Department of Physics, Kyungnam University,
Masan, 631-701, Korea.}
\date{\today}
\maketitle

\begin{abstract}
Following the Regge-Wheeler algorithm, we derive a radial equation 
for the brane-localized graviton absorbed/emitted by the $(4+n)$-dimensional
Schwarzschild black hole. Making use of this equation the absorption and 
emission spectra of the brane-localized graviton are computed numerically.
Existence of the extra dimensions generally suppresses the absorption
rate and enhances the emission rate as other spin cases. The appearance of the
potential well, however, when $n > \sqrt{\ell (\ell + 1) -2} - 1$ in the 
effective potential makes the decreasing behavior of the total absorption
with increasing $n$ in the low-energy regime. 
The high-energy limit of the total absorption cross section seems to 
coincide with that of the brane-localized scalar cross section.
The increasing rate of the graviton emission is very large compared to 
those of other brane-localized fields. This fact indicates 
that the graviton emission can be
dominant one in the Hawking radiation of the higher-dimensional black holes
when $n$ is large.
\end{abstract}

\newpage
The most striking result of the modern brane-world scenarios\cite{bwsc1}
is the emergence of the TeV-scale gravity when the extra dimensions exist.
This fact opens the possibility to make the tiny black holes in the 
future high-energy colliders\cite{hec1} on condition that the large or 
warped extra dimensions exist. In this reason the absorption and 
emission problems for the higher-dimensional black holes were extensively 
explored recently.

The absorption and emission spectra for the brane-localized spin $0$,
$1/2$ and $1$ particles and the bulk scalar in the $(4+n)$-dimensional 
Schwarzschild phase 
were numerically examined in Ref\cite{kanti1}. The main motivation of 
Ref.\cite{kanti1} is to check the issue\cite{emp00} 
whether or not the black holes radiate mainly on the brane. The result of 
the numerical calculation strongly supports the argument made by 
Emparan, Horowitz and Myers(EHM), {\it i.e.} the emission on the brane
is dominant compared to that off the brane.

In Ref.\cite{jung05-1} different numerical method was applied to examine the 
absorption and emission spectra of the brane-localized and bulk scalars in 
the background of the higher-dimensional charged black hole. 
The numerical method
used in Ref.\cite{jung05-1} is an application of the 
quantum mechanical scattering 
method incorporated with an analytic continuation, which was used in 
Ref.\cite{sanc78} to compute the spectra of the massless and massive scalars 
in the $4d$ Schwarzschild background. The numerical result of 
Ref.\cite{jung05-1} also supports the argument made by EHM if the number of 
the extra dimensions is not too large.

However, it was argued in Ref.\cite{frol1} that the argument of EHM should be 
examined carefully in the rotating black hole background 
because there is an another 
important factor called superradiance\cite{superr1} if the black holes 
have the angular momenta. Especially, in Ref.\cite{frol1} the existence of the 
superradiant modes was explicitly proven in the scattering of the bulk 
scalar when the spacetime background is a $5$-dimensional rotating 
black hole with two angular momentum parameters. The general criteria for 
the existence of the superradiant modes were derived in Ref.\cite{jung05-2} 
when the bulk scalar, electromagnetic and gravitational waves are absorbed
by the higher-dimensional rotating black hole with arbitrarily multiple
angular momentum parameters. 

The Hawking radiation on the brane is also examined when the bulk black hole
has an angular momentum\cite{ida02}. Numerical calculation supports the 
fact that the superradiance modes also exist in the brane emission. Thus the
brane emission and bulk emission were compared with each 
other in Ref.\cite{jung05-3} when the scalar field interacts with a 
$5$-dimensional rotating black hole. It was shown that although 
the superradiance
modes exist in the bulk emission in the wide range of energy, the energy
amplification for the bulk scalar is extremely small compared to the energy
amplification for the brane scalar. As a result the effect of the 
superradiance modes is negligible and does not change the EHM's claim.

There is an another factor we should take into account carefully in the 
Hawking radiation of the higher-dimensional black holes. This is an higher
spin effect like a graviton emission. 
Since the graviton is not generally 
localized on the brane unlike the standard model particles, the argument of the
EHM should be carefully checked in the graviton emission.
Recently, the bulk graviton emission was 
explored in Ref.\cite{corn05} when the background is an higher-dimensional
Schwarzschild black hole. In this paper we would like to examine the 
absorption and emission problems for the brane-localized graviton in the
background of the $(4+n)$-dimensional Schwarzschild black hole. 

We start with a metric induced by a $(4 + n)$-dimensional Schwarzschild 
spacetime\cite{tang63} 
\begin{equation}
\label{metric1}
ds_{\mbox{ind}}^2 = - h(r) dt^2 + h^{-1}(r) dr^2 + 
r^2 (d \theta^2 + \sin^2 \theta
d\phi^2)
\end{equation}
where $h(r) = 1 - (r_H / r)^{n+1}$. Of course, $r_H$ is a parameter which 
denotes the radius of the event horizon and the Hawking temperature is given
by $T_H = (n+1) / 4\pi r_H$. To derive an equation which 
governs the gravitational
fluctuation we should change the metric itself. Following the Regge-Wheeler
algorithm\cite{reg57} we introduce the gravitational 
fluctuation by adding the 
metric component
\begin{equation}
\label{metric2}
\delta s^2 = \left[H_0(r) dt d\phi + H_1(r) dr d\phi\right] e^{i \omega t}
\sin \theta \frac{d P_{\ell}}{d \theta} (\cos \theta)
\end{equation}
to the original metric $ds_{\mbox{ind}}^2$\footnote{This metric change 
corresponds to the odd parity gravitational perturbation. 
The even parity\cite{zer70}
perturbation leads a more complicate radial equation and will not be discussed
here.}. Of course, we should 
assume $H_0, H_1 << 1$ for the linearizarion. 

Now, we would like to discuss how to derive the fluctuation equation. Since
the metric (\ref{metric1}) is an usual Schwarzschild metric when $n=0$, it is 
a vacuum solution of the Einstein equation. For the $4d$ case, therefore, the
fluctuation equation can be expressed as $\delta R_{\mu \nu} = 0$, where
$R_{\mu \nu}$ is a Ricci tensor. For $n \neq 0$ case, however, the metric
(\ref{metric1}) is not a vaccum solution. Thus, we should assume that the 
metric (\ref{metric1}) is a non-vacuum solution of the $4d$ Einstein 
equation, {\it i.e.} ${\cal E}_{\mu \nu} = T_{\mu \nu}$ where
${\cal E}_{\mu \nu}$ and $T_{\mu \nu}$ are Einstein tensor and the 
non-vanishing energy-momentum tensor. Adding the metric perturbation
(\ref{metric2}) to the original metric (\ref{metric1}) the Einstein equation
would be changed into $\delta {\cal E}_{\mu \nu} = \delta T_{\mu \nu}$.
The main problem in the derivation of the fluctuation equation is the fact
that we do not know the nature of the matter because the non-vanishing 
energy-momentum tensor is not originated from the real matter on the 
brane but appears effectively in the course of the projection of the bulk
metric to the brane. Thus the difficulty is how to derive $\delta T_{\mu \nu}$
without knowing the real nature of the matter.

Firstly, we should note that the linear order perturbation yields only
non-vanishing $\delta {\cal E}_{t \phi}$, $\delta {\cal E}_{r, \phi}$ and
$\delta {\cal E}_{\theta \phi}$ in the gravity side. Thus it is reasonable
to assume that the non-vanishing components of $\delta T_{\mu \nu}$ are 
$\delta T_{t \phi}$, $\delta T_{r \phi}$ and $\delta T_{\theta \phi}$. The 
corresponding components of the Einstein equation are
\begin{eqnarray}
\label{linear1}
& & -\frac{1}{4} h(r) \frac{d^2 H_0}{d r^2} + 
\left( \frac{\lambda + h}{2 r^2} + \frac{1}{2 r} \frac{d h}{d r} + 
\frac{1}{4} \frac{d^2 h}{d r^2} \right) H_0 + \frac{i \omega}{4} h(r)
\left( \frac{d H_1}{d r} + \frac{2}{r} H_1 \right) = \delta \Theta_{t \phi}
                                                               \\   \nonumber
& &-\frac{i \omega}{4} h^{-1}(r) 
\left(\frac{d H_0}{d r} - \frac{2}{r} H_0 \right) + 
\left[ \frac{\lambda}{2 r^2} - \frac{\omega^2}{4} h^{-1}(r) +
      \frac{1}{2 r} \frac{d h}{d r} + \frac{1}{4} \frac{d^2h}{d r^2} \right]
H_1 = \delta \Theta_{r \phi}
                                                    \\    \nonumber
& &\hspace{3.0cm} 
i \omega h^{-1}(r) H_0 - h(r) \frac{d H_1}{d r} - \frac{d h}{d r} H_1 = 
      4 \delta \Theta_{\theta \phi}
\end{eqnarray}
where $\lambda = (\ell + 2) (\ell - 1) / 2$ and 
\begin{eqnarray}
\label{energy-momentum}
& & \delta T_{t \phi} = \delta \Theta_{t \phi}(r) e^{i \omega t} \sin \theta
\frac{d P_{\ell}}{d \theta} (\cos \theta)
                                                   \\   \nonumber
& & \delta T_{r \phi} = \delta \Theta_{r \phi}(r) e^{i \omega t} \sin \theta
\frac{d P_{\ell}}{d \theta} (\cos \theta)
                                                   \\    \nonumber
& & \delta T_{\theta \phi} = \delta \Theta_{\theta \phi}(r) e^{i \omega t}
\left( \cos \theta \frac{d}{d \theta} - \sin \theta \frac{d^2}{d \theta^2}
         \right) P_{\ell} (\cos \theta).
\end{eqnarray}

The conservation of the energy-momentum tensor yields a constraint
\begin{eqnarray}
\label{conservation}
& &\frac{i \omega}{r^2} h^{-1}(r) \delta \Theta_{t \phi} - \frac{1}{r^2}
h(r) \left[\partial_r \delta \Theta_{r \phi} + \left(h^{-1}(r) \frac{d h}{d r}
                                                     + \frac{2}{r} \right)
                                           \delta \Theta_{r \phi} \right]
- \frac{2 \lambda}{r^4} \delta \Theta_{\theta \phi}
                                                     \\   \nonumber
& &= \frac{i \omega}{2 r^3} h^{-1}(r) 
     \left(\frac{d h}{d r} + \frac{r}{2} \frac{d^2 h}{d r^2} \right) H_0
   - \frac{1}{2 r^3} h(r) \left(\frac{d h}{d r} + 
                  \frac{r}{2} \frac{d^2 h}{d r^2} \right) \frac{d H_1}{d r}
                                                     \\   \nonumber
& &\hspace{0.5cm}- \frac{1}{r^3}
\left[ \frac{1}{2} \frac{d h}{d r} \left(\frac{d h}{d r} + \frac{r}{2}
              \frac{d^2 h}{d r^2} \right) + h(r)
       \left( \frac{1}{2 r} \frac{d h}{d r} + \frac{d^2 h}{d r^2} + \frac{r}{4}
              \frac{d^3 h}{d r^3} \right) \right] H_1.
\end{eqnarray}
It is easy to show that as expected the rhs of Eq.(\ref{conservation}) is zero
when $n=0$. Another constraint may be derived from the fact that only two of 
three equations in Eq.(\ref{linear1}) should be linearly independent. However,
it does not yield a new constraint because it is possible to derive the 
first equation of (\ref{linear1}) from the remaining ones if 
(\ref{conservation}) is used. Final requirement is that the effective potential
derived from Eq.(\ref{linear1}) should coincide with the usual Regge-Wheeler
potential when $n=0$. This restriction gives a constraint
\begin{equation}
\label{constraint2}
\lim_{n \rightarrow 0} \delta \Theta_{t \phi} = 
\lim_{n \rightarrow 0} \delta \Theta_{r \phi} = 
\lim_{n \rightarrow 0} \delta \Theta_{\theta \phi} = 0.
\end{equation}
As far as we know, we cannot make any more constraint 
for the energy-momentum tensor from the general principle.
Of course the constraints (\ref{conservation}) and (\ref{constraint2}) are
not sufficient to fix $\delta \Theta_{t \phi}$, $\delta \Theta_{r \phi}$
and $\delta \Theta_{\theta \phi}$ completely. Thus we cannot determine the 
effective potential due to the lack of knowledge on the matter content.
Thus what we can do is to choose $\delta \Theta_{\mu \nu}$ consistently
with Eq.(\ref{conservation}) and (\ref{constraint2}).

We will choose 
$\delta \Theta_{r \phi} = \delta \Theta_{\theta \phi} = 0$ with a hope that the 
detailed shape of the effective potential does not change the graviton 
emissivity drastically\footnote{Of course, there are many other choices 
different from $\delta \Theta_{r \phi} = \delta \Theta_{\theta \phi} = 0$  
which satisfy Eq.(\ref{conservation}) and Eq.(\ref{constraint2})
simultaneously. Furthermore, the different choice may yield different and 
probably more complicate effective potential. The reason we choose
$\delta \Theta_{r \phi} = \delta \Theta_{\theta \phi} = 0$ is not because
it is more physically reasonable than other choices but because it yields
an effective potential via simple procedure. Thus our choice is valid 
physically when this assumption is true.}.
Eliminating $H_0$ in Eq.(\ref{linear1}), one can derive a differential 
equation for solely $H_1$ in the form
\begin{eqnarray}
\label{linear2}
& &\frac{d^2 H_1}{d r^2} + 
\left( 3 h^{-1}(r) \frac{d h}{d r} - \frac{2}{r} \right) \frac{d H_1}{d r}
                                                     \\   \nonumber
& & \hspace{1.0cm}
+ \left[ \left( \omega^2 + \left(\frac{d h}{d r} \right)^2 \right) h^{-2}(r)
        - \left(\frac{\ell (\ell + 1) - 2}{r^2} + \frac{4}{r} \frac{d h}{d r}
        \right) h^{-1}(r) \right] H_1 = 0.
\end{eqnarray}

Now, it is convenient to introduce a function $R(r)$ defined as 
$H_1 \equiv r^2 h^{-1}(r) R$. Then the fluctuation equation (\ref{linear2})
reduces to 
\begin{equation}
\label{linear3}
\frac{h(r)}{r^2} \frac{d}{d r}
\left( r^2 h(r) \frac{d R}{d r} \right) + 
\left[ \omega^2 - h(r) 
\left\{\frac{\ell (\ell + 1)}{r^2} - \epsilon
\left( \frac{2}{r^2} (1 - h(r)) - \frac{d^2 h}{d r^2} \right) \right\} \right]
                                 R.
\end{equation}
In Eq.(\ref{linear3}) we introduced a parameter $\epsilon$. When 
$\epsilon = 1$, Eq.(\ref{linear3}) corresponds to the gravitational 
fluctuation and $\epsilon = 0$ corresponds to the scalar fluctuation.
Thus one can treat the scalar and gravitational fluctuations in an unified way.
We used $\epsilon$ to confirm our numerical calculation which will be 
carried out later. Putting $\epsilon = 0$, we checked our numerical 
calculation reproduces the absorption and emission spectra of the 
brane-localized scalar calculated in Ref.\cite{kanti1,jung05-1}.

Defining again $R = \Psi / r$, Eq.(\ref{linear3}) is transformed into the 
Schr\"{o}dinger-like equation
\begin{equation}
\label{schro1}
\left(h(r) \frac{d}{d r}\right)^2 \Psi + 
\left[\omega^2 - V_{\mbox{eff}}(r) \right] \Psi = 0
\end{equation}
where the effective potential is 
\begin{equation}
\label{ef-poten1}
V_{\mbox{eff}}(r) = h(r)
\left[ \frac{\ell (\ell + 1)}{r^2} + \frac{1}{r} \frac{d h}{d r}
      - \epsilon \left( \frac{2}{r^2} \left( 1 - h(r) \right) - 
                      \frac{d^2 h}{d r^2} \right)             \right].
\end{equation}

It is worthwhile noting that until this stage we have not used the explicit
expression of $h(r)$. Thus we can use the equations derived above for the 
gravitational fluctuation of any $4d$-induced spherically symmetric metric.
Using an explicit expression of $h(r)$, one can easily show
\begin{equation}
\label{ef-poten2}
V_{\mbox{eff}}(r) = h(r)
\left[ \frac{\ell (\ell + 1)}{r^2} - \frac{\sigma_n r_H^{n+1}}{r^{n+3}}
                                                  \right]
\end{equation}
where $\sigma_n = \epsilon [(n+1) (n+2) + 2] - (n+1)$. 
When $n=0$, the effective potential (\ref{ef-poten2}) reduces to the usual
Regge-Wheeler potential.
Defining a 
``tortoise'' coordinate $z$ as $dz / dr = h^{-1}(r)$, the wave equation
(\ref{schro1}) simply reduces to
\begin{equation}
\label{schro2}
\frac{d^2 \Psi}{d z^2} + 
\left[ \omega^2 - V_{\mbox{eff}}(r) \right] \Psi = 0.
\end{equation}

\begin{figure}[ht!]
\begin{center}
\epsfysize=6.5cm \epsfbox{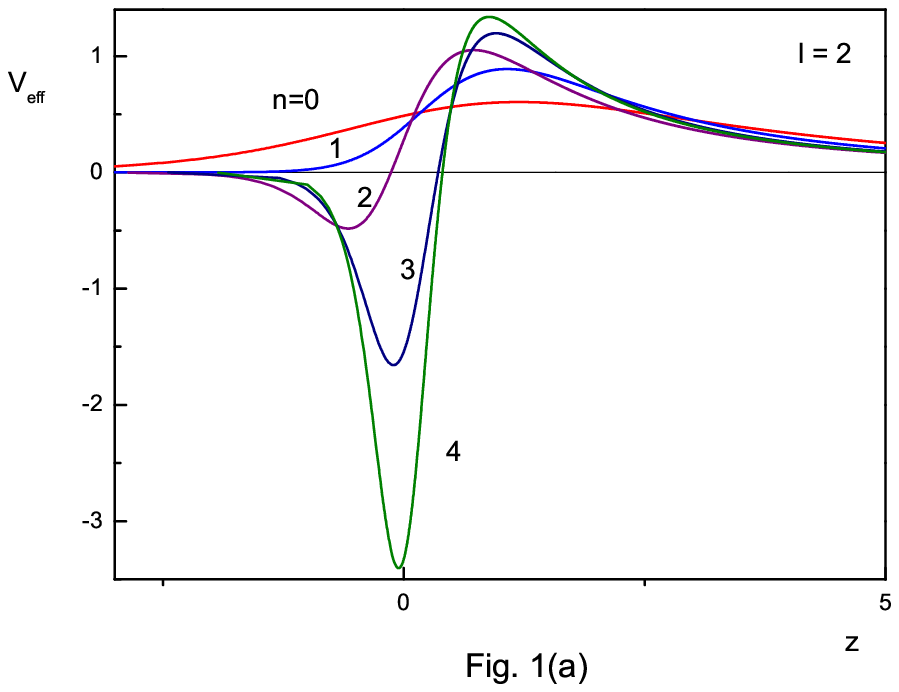}
\epsfysize=6.5cm \epsfbox{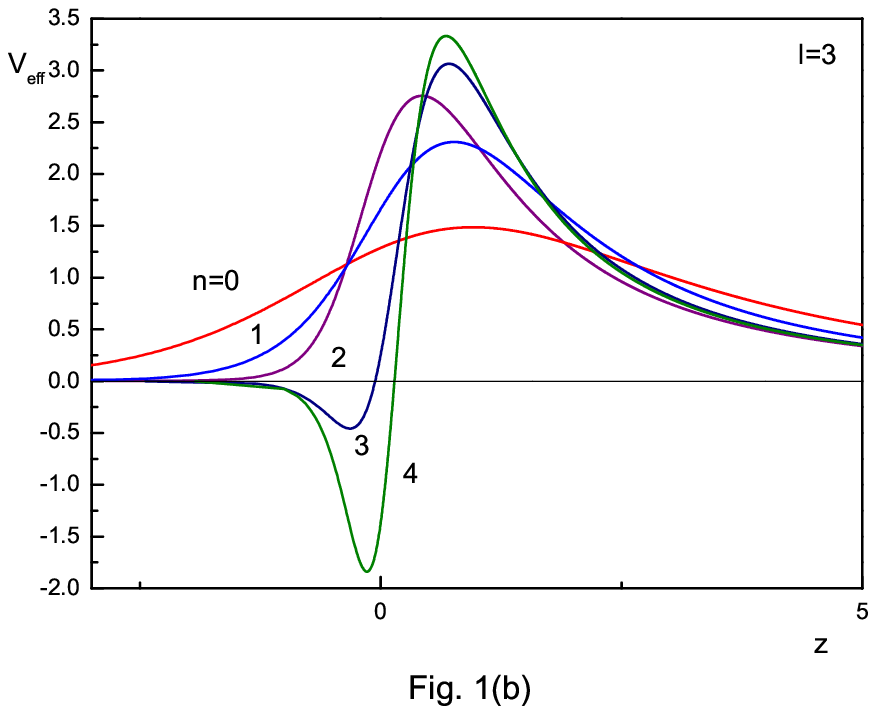}
\caption[fig1]{The $z$-dependence of the effective potential for the 
brane-localized graviton. Fig. 1(a) and (b) correspond to 
$\ell = 2$(the lowest mode) and 
$\ell = 3$ respectively. For fixed $\ell$ each potential with $n < \ell$ has
a barrier whose height increases with increasing $n$. When $n \geq \ell$, each
potential has a barrier in the side of the asymptotic region and a well in the
opposite side. The emergence of this potential well in the effective potential
seems to affect the low-energy behavior of the cross sections.} 
\end{center}
\end{figure}

The effective potential for the gravitational fluctuation, 
{\it i.e.} $\epsilon = 1$,is 
plotted as a function of $z$ in Fig. 1 when $\ell = 2$ (Fig. 1(a)) and 
$\ell = 3$ (Fig. 1(b)). For the lowest mode $\ell = 2$ each potential with
$n < \ell$ has a barrier whose height increases with increasing $n$. Thus
the existence of the extra dimensions should suppress the absorption rate.
Each potential with\footnote{The exact condition for the 
appearance of the potential well can be derived from the effective 
potential as $n > \sqrt{\ell (\ell + 1) - 2} - 1$.} $n \geq \ell$
has a barrier in the side of the asymptotic
region and a well in the opposite side. When $n$ increases, the barrier and
well become higher and deeper respectively. While, therefore, the barrier 
tends to decrease the absorption rate with increasing $n$, the well tends to 
enhance it. Thus, the absorption cross section should be determined by the 
competition between the barrier and the well. The effective potential with 
$\ell = 3$ exhibits a similar behavior as Fig. 2(b) shows.

The absorption cross section for $\ell = 2$ and $3$ is plotted in Fig. 2 when
$n = 0$, $1$, and $2$. Fig. 2 indicates that overall, the existence of the
extra dimensions suppresses the absorption rate as other fields. This means 
that the potential barrier is more important factor than the potential well
in determining the absorption rate. However, Fig. 2(a) shows that the 
cross section for $n=2$ is larger than that for $n=0$ or $1$ in the 
low-$\omega$ region. This means that the potential well plays more 
important role in this region.

\begin{figure}[ht!]
\begin{center}
\epsfysize=6.3cm \epsfbox{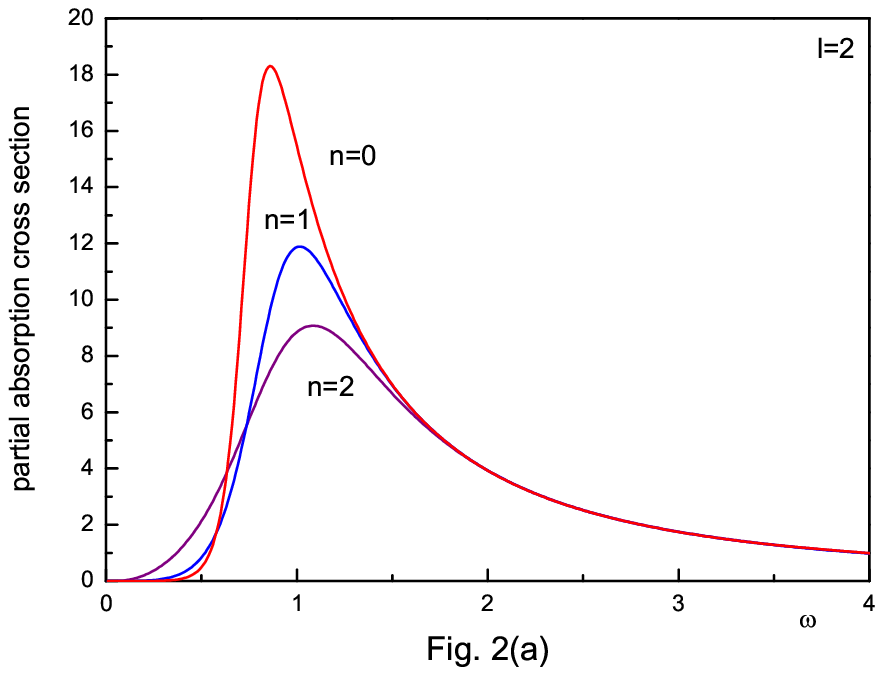}
\epsfysize=6.3cm \epsfbox{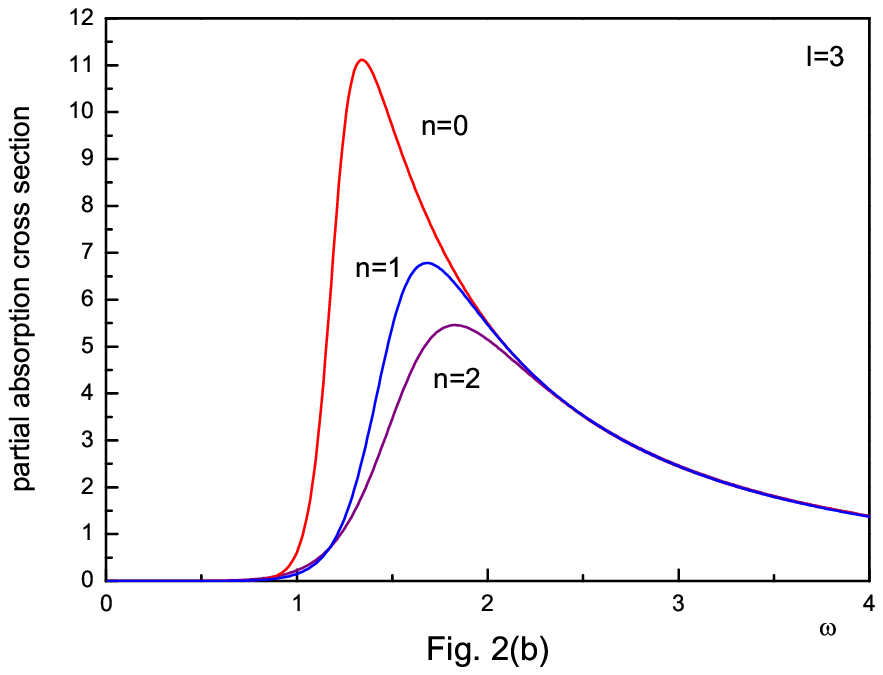}
\caption[fig2]{The $\omega$-dependence of the partial absorption cross sections
for $\ell = 2$(Fig. 2(a)) and $\ell = 3$(Fig. 2(b)). Overall, the existence 
of the extra dimensions suppresses the absorption rate as in the cases of other
fields. However, the reversion of the magnitude in the low-energy regime 
of Fig. 2(a) seems to be due to the emergence of the potential well in the
effective potential.} 
\end{center}
\end{figure}

Now we would like to explain how to compute the physical quantities related
to the absorption and emission of the brane-localized graviton. Since the
numerical method used in this paper was extensively used 
in Ref.\cite{jung05-1,sanc78,jung05-3},
we will just comment it schematically. Using the dimensionless parameters
$x \equiv \omega r$ and $x_H \equiv \omega r_H$, the radial equation
(\ref{linear3}) can be written as 
\begin{eqnarray}
\label{radial1}
& &\hspace{2.0cm} x^2 (x^{n+1} - x_H^{n+1})^2 \frac{d^2 R}{d x^2} + 
x (x^{n+1} - x_H^{n+1})
\left[2 x^{n+1} + (n-1) x_H^{n+1}\right]  \frac{d R}{d x}  \\  \nonumber
& & 
+ \left[x^{2 n + 4} - \ell (\ell + 1) x^{n+1} (x^{n+1} - x_H^{n+1})  
        + \epsilon \left\{(n+1) (n+2) + 2\right\} x_H^{n+1}
              (x^{n+1} - x_H^{n+1}) \right] R = 0.
\end{eqnarray}
Due to the real nature of Eq.(\ref{radial1}), $R^{\ast}$ should be a solution
if $R$ is a solution of Eq.(\ref{radial1}). The Wronskian between them is 
\begin{equation}
\label{wron1}
W[R^{\ast}, R]_x \equiv R^{\ast} \frac{d R}{d x} - R \frac{d R^{\ast}}{d x}
= \frac{{\cal{C}} x^{n-1}}{x^{n+1} - x_H^{n+1}}
\end{equation}
where ${\cal{C}}$ is an integration constant. Since $x = x_H$ is a regular
singular point of the differential equation (\ref{radial1}), we can solve 
Eq.(\ref{radial1}) to obtain a solution which is convergent in the near-horizon
regime as following:
\begin{equation}
\label{near-hor1}
{\cal{G}}_{n,\ell} (x, x_H) = e^{\lambda_n \ln |x - x_H|}
\sum_{N=0}^{\infty} d_{\ell, N} (x - x_H)^N
\end{equation}
where $\lambda_n = -i x_H / (n + 1)$. Of course, the recursion relation can
be directly derived by inserting Eq.(\ref{near-hor1}) into the radial
equation (\ref{radial1}). The Wronskian between ${\cal{G}}^{\ast}$ and 
${\cal{G}}$ is same with Eq.(\ref{wron1}) with 
${\cal{C}} = -2 i |d_{\ell,0}|^2 x_H^2$. In Ref.\cite{sanc78} 
the relation between
$|d_{\ell,0}|^2$ and the partial transmission coefficient 
${\cal{T}}_{n,\ell}$ 
\begin{equation}
\label{rela1}
|d_{\ell,0}|^2 = \frac{(\ell + 1/2)^2}{x_H^2} {\cal{T}}_{n,\ell}
\end{equation}
is derived by making use of the quantum mechanical scattering theory.

The ingoing and outgoing solutions which are convergent in the asymptotic
regime can be derived directly also as series expressions in the form
\begin{equation}
\label{asymp1}
{\cal{F}}_{n,\ell(\pm)}(x, x_H) = (\pm i)^{\ell + 1} e^{\mp i x} 
(x - x_H)^{\pm \lambda_n} \sum_{N=0}^{\infty} \tau_{N(\pm)} x^{-(N+1)}.
\end{equation}
The Wronskian between ${\cal{F}}_{(+)}$ and ${\cal{F}}_{(-)}$ is same with
Eq.(\ref{wron1}) with ${\cal{C}} = 2 i$.

Since the asymptotic solution is a linear combination of the ingoing and 
outgoing solutions, one can assume that the scattering solution  
which is convergent at the asymptotic region is 
\begin{equation}
\label{asymp2}
{\cal{R}}_{n,\ell}(x, x_H) = f_{n,\ell}^{(-)}(x_H) {\cal{F}}_{n,\ell(+)}
(x, x_H) + f_{n,\ell}^{(+)}(x_H) {\cal{F}}_{n,\ell(-)}
(x, x_H)
\end{equation}
where the coefficients $f_{n, \ell}^{(\pm)}$ are called the jost functions.
It should be noted that if we know ${\cal{R}}$, the jost functions can be 
computed by the Wronskian as following:
\begin{equation}
\label{wron2}
f_{n, \ell}^{(\pm)}(x_H) = \pm \frac{x^{n+1} - x_H^{n+1}}{2 i x^{n-1}}
W[{\cal{F}}_{n,\ell(\pm)}, {\cal{R}}_{n,\ell}]_x. 
\end{equation}
In Ref.\cite{sanc78} the relation
between the jost functions and the coefficient $d_{\ell,0}$ is also derived
in the form
\begin{equation}
\label{rela2}
d_{\ell,0} = \frac{\ell + 1/2}{f_{n, \ell}^{(-)}}.
\end{equation}

Combining Eq.(\ref{rela1}) and (\ref{rela2}) the partial absorption cross
section can be written as 
\begin{equation}
\label{abss1}
\sigma_{\ell}^{BR} \equiv \frac{\pi}{\omega^2} (2\ell + 1) 
{\cal{T}}_{\ell} = 
\frac{\pi (2 \ell + 1) r_H^2}{|f_{n,\ell}^{(-)}(x_H)|^2}
\end{equation}
where the superscript denotes the brane-localized graviton.

The emission spectrum, {\it i.e.} the energy emitted per unit time, is given 
by the Hawking formula
\begin{equation}
\label{emis1}
\Gamma^{BR} = \frac{\omega^3 \sigma_{abs}^{BR}}{2 \pi^2 (e^{\omega / T_H} - 1)}
d \omega
\end{equation}
where $\sigma_{abs}^{BR}$ is a total absorption cross section. Thus the 
absorption and emission spectra can be completely computed if we know the jost
functions. In Ref.\cite{jung05-1,sanc78,jung05-3}
the function ${\cal{R}}$ is computed numerically
from the near-horizon solution (\ref{near-hor1}) by increasing the convergence
region  via the analytic continuation. Then Eq.(\ref{wron2}) enables us to 
compute the jost functions numerically. This is an overall algorithm of our
numerical technique.

\begin{figure}[ht!]
\begin{center}
\epsfysize=8.0cm \epsfbox{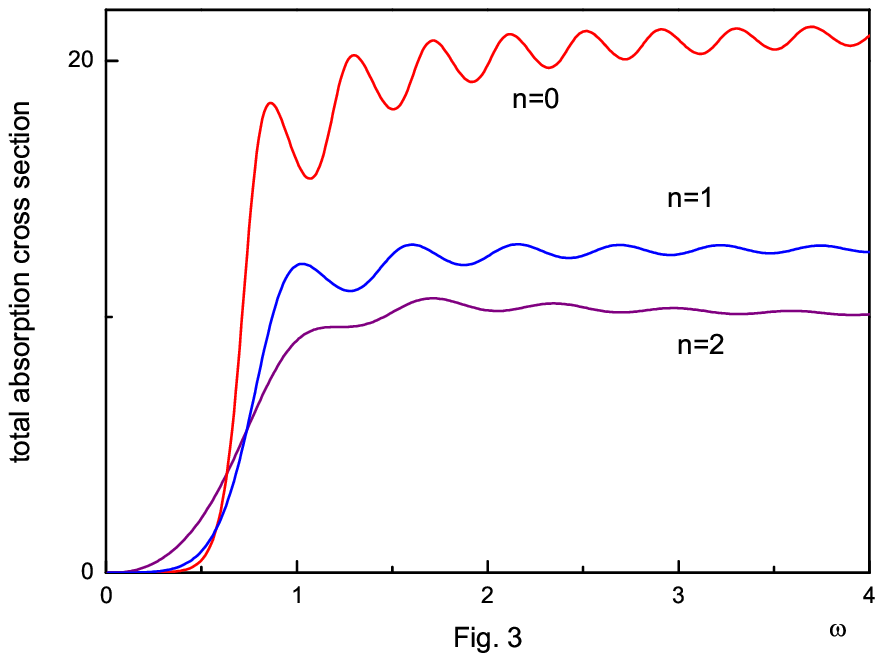}
\caption[fig3]{The $\omega$-dependence of the total absorption cross sections 
for the brane-localized graviton. Overall, the existence of the extra 
dimensions suppresses the total absorption cross sections. The low-energy 
behavior of this figure also implies the crucial effect of the potential
well emerged when $n \geq \ell$.} 
\end{center}
\end{figure}

The total absorption cross section is plotted in Fig. 3 when $n=0$, $1$ and 
$2$. As in the case of the scalar field absorption the existence of the 
extra dimensions in general suppresses the absorption rate. 
This indicates that the potential barrier is more important factor in
determining the absorption cross section. However, the low-$\omega$ region
of Fig. 3 seems to indicate that the emergence of the potential well
when $n \geq \ell$ crucially changes the order of magnitude.
The wiggly pattern
indicates that each partial absorption cross section has a peak point
at different $\omega$. The high-energy limits of the total cross sections seem
to coincide with the scalar case, {\it i.e.}
\begin{equation}
\label{high1}
\lim_{\omega \rightarrow \infty} \sigma_{abs}^{BR} = 
\left(\frac{n+3}{2}\right)^{2 / (n+1)} \frac{n+3}{n+1} \pi r_H^2.
\end{equation}

\begin{figure}[ht!]
\begin{center}
\epsfysize=8.0cm \epsfbox{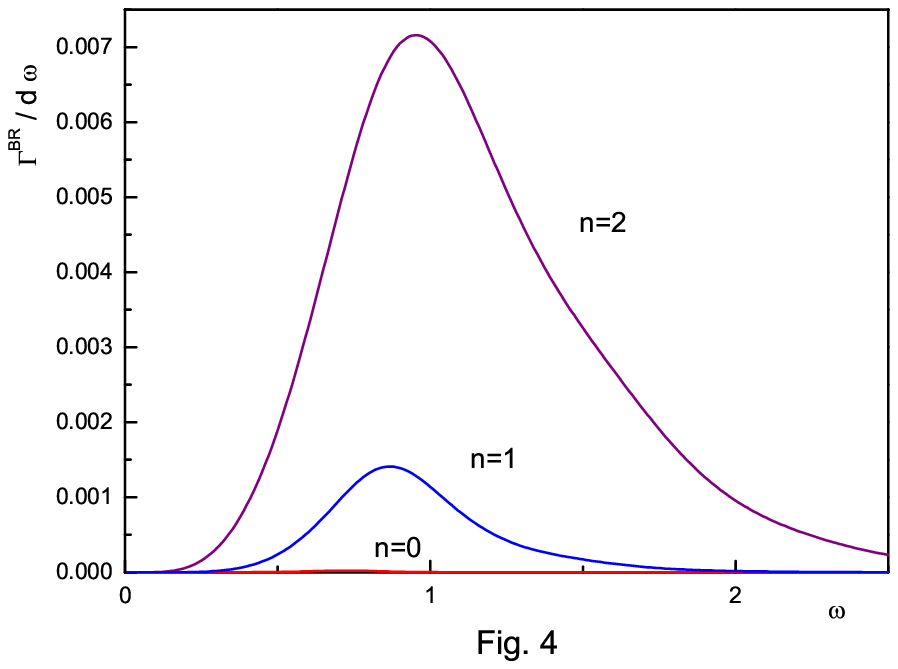}
\caption[fig4]{The $\omega$-dependence of the emission rate when $n=0$, $1$,
and $2$. As in the cases of other fields the existence of the extra 
dimensions generally enhances the emissivities. The emission spectrum for 
$n=0$ cannot be seen clearly because it is too small compared to other cases.
The drastic change of the emission rate indicates that the graviton emission 
can be a dominant one in the Hawking radiation of the higher-dimensional 
black holes for large $n$.} 
\end{center}
\end{figure}

The emission spectrum of the graviton is plotted in Fig. 4. As in the case 
of the scalar emission the existence of the extra dimensions enhances the 
emission rate. This indicates that the Planck factor is most important
in determining the emission rate. 
For $n=0$ the emission spectrum cannot be seen clearly in Fig. 4 because it is 
too small compared to the other cases. The drastic increase of the emission
rate implies that the Hawking radiation of the higher spin becomes more 
and more important for large $n$.
\vspace{1.0cm}
\begin{center}

\begin{tabular}{l|l|l|r} \hline
$ $           &   $n=0$   &     $n=1$     &     $n=2$     \\  \hline \hline
graviton    &   $7.67 \times 10^{-6}$   &  $8.06 \times 10^{-4}$  &  
                                             $6.70 \times 10^{-3}$      \\
scalar      &   $2.98 \times 10^{-4}$     &  $2.66 \times 10^{-3}$    &  
                                             $1.07 \times 10^{-2}$  \\ 
graviton/scalar  &  $0.026$   &  $0.303$   &   $0.624$         \\  \hline
\end{tabular}

\vspace{0.1cm}
\large{Table I} 
\end{center}
\vspace{1.0cm}

The total emission rate $\int \Gamma^{BR}$ for graviton and scalar 
is summarized 
in Table I. Generally the total emission rate for the graviton is smaller 
than that for the scalar. But the ratio between them tends to increase
dramatically with
increasing $n$. Thus we can conjecture that for large $n$ the graviton 
radiation becomes dominant in the Hawking radiation of the higher-dimensional
black holes.

In this paper we studied the absorption and emission of the brane-localized
graviton in the context of the brane-world scenarios.  
The existence of the extra dimensions in general tends to suppress the 
total absorption cross section as the cases of other fields. In the low-energy
region, however, the order of the magnitude is reversed (see Fig. 3). 
This seems to be  
due to the emergence of the well in the effective potential. 
The high-energy limits 
of the total cross sections for
the brane-localized gravitons seem to coincide with those for the 
brane-localized massless scalars. 

As in the cases of other fields the existence of the extra dimensions
generally enhances the emission rate. 
This fact indicates that the Planck factor is most dominant term in determining 
the emissivities.
However, the increasing rate of the brane-localized graviton is very large
compared to those of other fields. To show this explicitly, the increasing rates
of the brane-localized graviton and scalar is given in Table I. This rapid
increase of the emissivity indicates that the graviton emission should be 
dominant radiation in the Hawking radiation of the higher-dimensional 
black holes when $n$ is large.

It is straightforward to extend our calculation to the higher-dimensional
charged black hole background by choosing
\begin{equation}
\label{charged1}
h(r) = \left[ 1 - \left( \frac{r_+}{r} \right)^{n+1} \right]
      \left[ 1 - \left( \frac{r_-}{r} \right)^{n+1} \right] 
\end{equation}
where $r_{\pm}$ correspond to the outer and inner horizons. Since we have not 
used the explicit expression of $h(r)$ when deriving Eq.(\ref{linear3}), 
we can use the radial equation (\ref{linear3}) with only choosing 
different $h(r)$. The introduction of the inner horizon parameter $r_-$ 
changes the Hawking temperature into
\begin{equation}
\label{charged2}
T_H = \frac{n + 1}{4 \pi r_+^{n+2}} \left(r_+^{n+1} - r_-^{n+1} \right).
\end{equation}
When, therefore, $r_-$ increases, the Hawking temperature becomes lower, 
which should enhance the absorptivity and reduce the emission rate. Since 
the introduction of the extra dimensions generally suppresses the absorptivity
and enhances the emission rate, the exact absorption and emission spectra
for the charged black hole case may be determined by the competition between 
$r_-$ and $n$.

Recently, in Ref.\cite{corn05} the absorption rate for the bulk graviton is 
studied in the Schwarzschild background. The fluctuations for the bulk 
graviton have three different modes, {\it i.e.} scalar, vector, and tensor
modes. It is of greatly interest to apply our method to compare the emission
rate for the brane-localized graviton with that for the bulk graviton.

\vspace{1cm}

{\bf Acknowledgement}:  
This work was supported by the Kyungnam University
Research Fund, 2006.


\begin{thebibliography}{99}
\bibitem{bwsc1} N. Arkani-Hamed, S. Dimopoulos and G. Dvali,
{\it The Hierarchy Problem and New Dimensions at a Millimeter}, 
Phys. Lett. {\bf B429} (1998) 263 [hep-ph/9803315]; 
L. Antoniadis, N. Arkani-Hamed, S. Dimopoulos and G. Dvali,
{\it New Dimensions at a Millimeter to a Fermi and Superstrings at a 
TeV}, Phys. Lett. {\bf B436} (1998) 257 [hep-ph/9804398]; 
L. Randall and R. Sundrum, {\it A Large Mass Hierarchy from a 
Small Extra Dimension}, 
Phys. Rev. Lett. {\bf 83} (1999) 3370 [hep-ph/9905221]; 
L. Randall and R. Sundrum, {\it An Alternative to 
Compactification}, Phys. Rev. Lett. {\bf 83} (1999) 4690 [hep-th/9906064].
\bibitem{hec1} S. B. Giddings and T. Thomas, {\it High energy colliders 
as black hole factories: The end of short distance physics}, Phys. Rev. 
{\bf D65} (2002) 056010 [hep-ph/0106219];
S. Dimopoulos and G. Landsberg, {\it Black Holes at the 
Large Hadron Collider}, Phys. Rev. Lett. {\bf 87} (2001) 161602 
[hep-ph/0106295];
D. M. Eardley and S. B. Giddings, {\it Classical black hole 
production in high-energy collisions}, Phys. Rev. {\bf D66} (2002)
044011 [gr-qc/0201034];
D. Stojkovic, {\it Distinguishing between the small ADD and
RS black holes in accelerators}, Phys. Rev. Lett. {\bf 94} (2005)
011603 [hep-ph/0409124];
V. Cardoso, E. Berti and M. Cavagli\`{a}, {\it What we
(don't) know about black hole formation in high-energy collisions},
Class. Quant. Grav. {\bf 22} (2005) L61 [hep-ph/0505125].
\bibitem{kanti1} C. M. Harris and P. Kanti, {\it Hawking Radiation from a
$(4+n)$-dimensional Black Hole: Exact Results for the Schwarzschild Phase},
JHEP 0310 (2003) 014 [hep-ph/0309054];
P. kanti, {\it Black Holes in Theories with Large Extra 
Dimensions: a Review}, Int. J. Mod. Phys. {\bf A19} (2004) 4899
[hep-ph/0402168].
\bibitem{emp00}P. Argyres, S. Dimopoulos and J. March-Russell, 
{\it Black Holes and Sub-millimeter Dimensions}, Phys. Lett. 
{\bf B441} (1998) 96 [hep-th/9808138];
T. Banks and W. Fischler, {\it A Model for High Energy Scattering
in Quantum Gravity} [hep-th/9906038]; 
R. Emparan, G. T. Horowitz and R. C. Myers, {\it Black Holes
radiate mainly on the Brane}, Phys. Rev. Lett. {\bf 85} (2000) 499
[hep-th/0003118].
\bibitem{jung05-1} E. Jung and D. K. Park, {\it Absorption and Emission 
Spectra of an higher-dimensional Reissner-Nordstr\"{o}m black hole}, 
Nucl. Phys. {\bf B717} (2005) 272 [hep-th/0502002].
\bibitem{sanc78} N. Sanchez, {\it Absorption and emission spectra of a
Schwarzschild black hole}, Phys. Rev. {\bf D18} (1978) 1030;
E. Jung and D. K. Park, {\it Effect of Scalar Mass in the 
Absorption and Emission Spectra of Schwarzschild Black Hole}, Class. Quant.
Grav. {\bf 21} (2004) 3717 [hep-th/0403251].
\bibitem{frol1}V. Frolov and D. Stojkovi\'{c}, {\it Black hole radiation
in the brane world and the recoil effect}, Phys. Rev. {\bf D66} (2002)
084002 [hep-th/0206046];
V. Frolov and D. Stojkovi\'{c}, {\it Black Hole as a Point
Radiator and Recoil Effect on the Brane World}, Phys. Rev. Lett. 
{\bf 89} (2002) 151302 [hep-th/0208102];
V. Frolov and D. Stojkovi\'{c}, {\it Quantum radiation from 
a $5$-dimensional black hole}, Phys. Rev. {\bf D67} (2003) 084004
[gr-qc/0211055].
\bibitem{superr1} Y. B. Zel'dovich, {\it Generation of waves by a 
rotating body}, JETP Lett. {\bf 14} (1971) 180;
W. H. Press and S. A. Teukolsky, {\it Floating Orbits, 
Superradiant Scattering and the Black-hole Bomb}, Nature {\bf 238} (1972) 211;
A. A. Starobinskii, {\it Amplification of waves during 
reflection from a rotating black hole}, Sov. Phys. JETP {\bf 37} (1973)
28;
A. A. Starobinskii and S. M. Churilov, {\it Amplification
of electromagnetic and gravitational waves scattered by a rotating black
hole}, Sov. Phys. JETP {\bf 38} (1974) 1.
\bibitem{jung05-2}E. Jung, S. H. Kim and D. K. Park, {\it Condition for 
Superradiance in Higher-dimensional Rotating Black Holes}, Phys. Lett.
{\bf B615} (2005) 273
[hep-th/0503163];
E. Jung, S. H. Kim and D. K. Park, {\it Condition for the
Superradiance Modes in Higher-Dimensional Black Holes with Multiple
Angular Momentum Parameters}, Phys. Lett. {\bf B619} (2005) 347
[hep-th/0504139].
\bibitem{ida02} D. Ida, K. Oda and S. C. Park, {\it Rotating black holes
at future collider: Greybody factors for brane field}, Phys. Rev. 
{\bf D67} (2003) 064025 [hep-th/0212108];
C. M. Harris and P. Kanti, {\it Hawking Radiation from a
$(4+n)$-Dimensional Rotating Black Hole} [hep-th/0503010];
D. Ida, K. Oda and S. C. Park, {\it Rotating black holes 
at future colliders II : Anisotropic scalar field emission},
Phys. Rev. {\bf D71} (2005) 124039 [hep-th/0503052];
G. Duffy, C. Harris, P. Kanti and E. Winstanley, {\it Brane decay of a 
$(4+n)$-dimensional rotating black hole: spin-$0$ particle},
JHEP {\bf 0509} (2005) 049 [hep-th/0507274];
M. Casals, P. Kanti and E. Winstanley, {\it Brane Decay of a 
$(4+n)$-Dimensional Rotating Black Hole II: spin-$1$ particles}
[hep-th/0511163].
\bibitem{jung05-3} E. Jung and D. K. Park, {\it Bulk versus Brane in the 
Absorption and Emission: $5$D Rotating Black Hole Case}, 
Nucl. Phys. {\bf B731} (2005) 171 [hep-th/0506204].
\bibitem{corn05} A. S. Cornell, W. Naylor and M. Sasaki,
{\it Graviton emission from a higher-dimensional black hole}
[hep-th/0510009].
\bibitem{tang63} F. R. Tangherlini, {\it Schwarzschild Field in $n$
Dimensions and the Dimensionality of Space Problem},
Nuovo Cimento {\bf 27} (1963) 636;
R. C. Myers and M. J. Perry, {\it Black Holes in Higher
Dimensional Space-Times}, 
Ann. Phys. {\bf 172} 
(1986) 304.
\bibitem{reg57} T. Regge and J. A. Wheeler, {\it Stability of a 
Schwarzschild Singularity}, Phys. Rev. {\bf 108} (1957) 1063;
C. V. Vishveshwara, {\it Stability of the Schwarzschild 
Metric}, Phys. Rev. {\bf D 1} (1970) 2870;
L. A. Edelstein and C. V. Vishveshwara, {\it Differential
Equations for Perturbations on the Schwarzschild Metric}, Phys. Rev. 
{\bf D 1} (1970) 3514;
C. V. Vishveshwara, {\it Scattering of Gravitational 
Radiation by a Schwarzschild Black-hole}, Nature {\bf 227} (1970) 936.
\bibitem{zer70} F. J. Zerilli, {\it Effective Potential for even-parity
Regge-Wheeler Gravitational Perturbation Equations}, Phys. Rev. Lett.
{\bf 24} (1970) 737.
\end{thebibliography}
\end{document}